\PassOptionsToPackage{table}{xcolor}
\documentclass[fleqn,10pt]{wlscirep}
\usepackage[utf8]{inputenc}
\usepackage[T1]{fontenc}
\usepackage{graphicx}%
\usepackage{multirow}%
\usepackage{amsmath,amssymb,amsfonts}%
\usepackage{amsthm}%
\usepackage[mathcal]{euscript}%
\usepackage[title]{appendix}%
\usepackage{xcolor}%
\usepackage{textcomp}%
\usepackage{booktabs}%
\usepackage{algorithm}
\usepackage{algorithmicx}
\usepackage{algpseudocode}%
\usepackage{listings}%
\usepackage{caption}
\usepackage{subcaption}
\usepackage{multibib}
\newcites{supp}{References}
\usepackage{hyperref}
\usepackage{float}
\usepackage{multirow}
\usepackage{pifont}

\definecolor{mygray}{gray}{0.8}

\DeclareMathOperator*{\argmax}{arg\max}
\DeclareMathOperator*{\argmin}{arg\min}

\title{Group-specific discriminant analysis reveals statistically validated sex differences in lateralization of brain functional network}

\author[1,2,+]{Shuo Zhou}
\author[3,+]{Junhao Luo}
\author[3]{Yaya Jiang}
\author[1,2]{Haolin Wang}
\author[1,2,*]{Haiping Lu}
\author[3,4,5,*]{Gaolang Gong}
\affil[1]{Department of Computer Science, University of Sheffield, Sheffield, UK}
\affil[2]{Centre for Machine Intelligence, University of Sheffield, Sheffield, UK}
\affil[3]{State Key Laboratory of Cognitive Neuroscience and Learning \& IDG/McGovern Institute for Brain Research, Beijing Normal University, Beijing, China}
\affil[4]{Beijing Key Laboratory of Brain Imaging and Connectomics, Beijing Normal University, Beijing, China}
\affil[5]{Chinese Institute for Brain Research, Beijing, China}

\affil[*]{h.lu@sheffield.ac.uk; gaolang.gong@bnu.edu.cn}

\affil[+]{These authors contributed equally to this work}

\begin{abstract}
Lateralization is a fundamental feature of the human brain, where sex differences have been observed. Conventional studies in neuroscience on sex-specific lateralization are typically conducted on univariate statistical comparisons between male and female groups. However, these analyses often lack effective validation of group specificity. Here, we formulate modeling sex differences in lateralization of functional networks as a dual-classification problem, consisting of first-order classification for left vs. right functional networks and second-order classification for male vs. female models. To capture sex-specific patterns, we develop the Group-Specific Discriminant Analysis (GSDA) for first-order classification. The evaluation on two public neuroimaging datasets demonstrates the efficacy of GSDA in learning sex-specific models from functional networks, achieving a significant improvement in group specificity over baseline methods. The major sex differences are in the strength of lateralization and the interactions within and between lobes. The GSDA-based method is generic in nature and can be adapted to other group-specific analyses such as handedness-specific or disease-specific analyses. 
\end{abstract}

\begin{document}

\flushbottom
\maketitle

\thispagestyle{empty}


\section*{Introduction}

\noindent 
Human brains are functionally lateralized \cite{ojemann1989cortical,gazzaniga2000cerebral}. 
The asymmetries between left and right brain hemispheres are believed to reflect a complex interplay of evolutionary, hereditary, developmental, experiential, and pathological influences \cite{corballis2017evolution}. 
Researchers are developing insights on lateralization through psychological, pharmacological, and neuroscience investigations \cite{passingham2002anatomical,toga2003mapping}. One important understanding is that multiple factors influence human brain lateralization \cite{toga2003mapping}, with sex being one of the most representative \cite{clements2006sex,tomasi2012laterality,agcaoglu2015lateralization, reber2017sex, guadalupe2015asymmetry,hirnstein2013sex, plessen2014sex,sommer2008sex}. A popular viewpoint is that males have a more asymmetric brain organization while females have a more ``bilateral'' brain organization, which may result in the males' superior spatial skills and the females' superior verbal skills \cite{levy1972lateral,levy1978lateral}. 

Measurement of functional brain lateralization is valuable but challenging \cite{jansen2006assessment}. Direct approaches such as selectively modulating or suppressing cortical activities and circuits in a single hemisphere \cite{kinsbourne1977hemineglect} often pose a risk of inflicting harm on the human brains \cite{kolb2009fundamentals}. 
Over the last two decades, functional neuroimaging techniques have been widely used in neuroscience, offering a powerful and non-invasive approach to studying human brain lateralization \cite{fox2007spontaneous,logothetis2008we}. 
One popular technique is analyzing functional connectivity (FC) of the brain's resting-state, which is also known as the brain network or connectome \cite{sporns2005human}. This is usually derived from functional magnetic resonance imaging (fMRI) time series and considered an intrinsic ``fingerprint'' of the human brain \cite{passingham2002anatomical,smith2012temporally,finn2015functional}.
A previous study \cite{agcaoglu2015lateralization} reported sex differences in lateralization of resting-state networks, with more right-lateralized visual and default-mode network components for males and females, respectively. Several other networks also showed differences between males and females \cite{agcaoglu2015lateralization}.
Additionally, males and females have also demonstrated significant differences in homotopic functional connectivity of numerous regions \cite{zuo2010growing}. 

Studies on brain lateralization have largely focused on modeling asymmetry effects region-by-region \cite{toga2003mapping}. These lateralized brain regions are usually measured by the laterality index (LI) \cite{seghier2008laterality,tomasi2012laterality}, or identified through statistical univariate analysis comparing homologous regions \cite{liegeois2002direct,friston2011functional}.
However, using these conventional methods to search for sex-specific lateralization patterns is mostly limited to within-group analysis \cite{jacobsen2007gender,schwarz2011sex, kret2011men, ingalhalikar2014sex, cui2018individualized,zullo2020gene}. For example, to understand male-specific lateralization, analyses are performed on male and female data separately to label features significantly differing from female data as ``male-specific''. Moreover, these analyses often do not validate the models on unseen samples, and therefore the resulting patterns may not be truly group-specific. In addition, given the high similarity between male and female brains, the small size of statistical effects makes detecting sex differences in lateralization more difficult \cite{good2001cerebral}. As a result, the true specificity of lateralization may be overwhelmed by the similarities. Hence, effectively modeling and validating sex-specific lateralization remains challenging.

\begin{figure*}[t]
    \includegraphics[width=\linewidth]{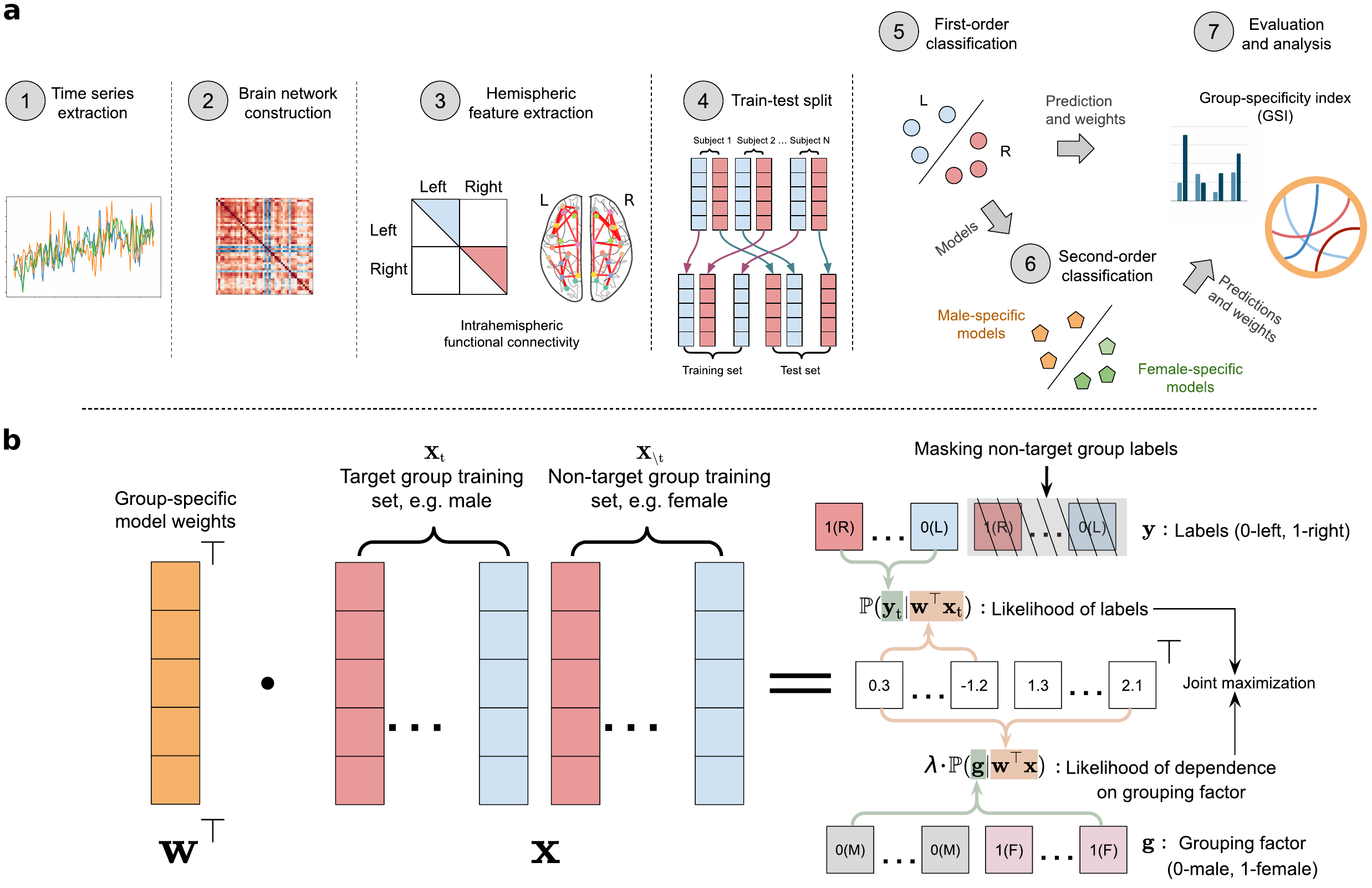}
    \caption{\textbf{a}, The proposed classification workflow for identifying sex-specific brain lateralization. \ding{172}--\ding{174}: Hemispheric features are extracted from the intrahemispheric brain functional network, which is constructed using resting-state functional MRI time series. \ding{175}\ding{176}: First-order classification learns the differences between the two hemispheres, where a group-specific discriminant analysis (GSDA) classifier is trained to classify left vs. right hemispheres for a target group. \ding{177}: Second-order classification trains a standard logistic regression for classifying the male- vs. female-specific models obtained from the first-order classification, to identify the weights that significantly contribute to the sex-specific predictions. \ding{178}: Evaluate the predictions and interpret the model weights. \textbf{b}, Group-specific discriminant analysis with the logistic loss (GSDA-Logit) for the first-order classification of left vs. right brain hemispheres. This model jointly maximizes the likelihood of labels for the target group (with non-target-group labels masked out) and the grouping factor dependence for both the target and non-target groups, where $^\top$ denotes the transpose of vectors, $\mathbf{x}$ denotes the input training samples, $\mathbf{x}_{\mathrm{t}}$ denotes target group training samples, $\mathbf{x}_{\setminus\mathrm{t}}$ denotes non-target group samples, and a hyperparameter $\lambda\ge 0$ controls the grouping factor dependence.
    }
\label{fig:gsda-method}
\end{figure*}

Here, we formulate the identification of sex differences in brain lateralization as a machine learning classification problem to address the aforementioned challenges. \textit{First}, to model sex differences in brain lateralization, we propose a dual classification workflow. This consists of a first-order classification of left vs. right brain hemispheres and a second-order classification of male- vs. female-specific models. 
The obtained first- and second-order model weights can be interpreted as lateralization strength and sex differences, respectively. The whole workflow is presented in Fig. \ref{fig:gsda-method}a. \textit{Second}, to learn group (sex)-specific models in the first-order classification, we propose a novel group-specific discriminant analysis (GSDA) algorithm (Fig. \ref{fig:gsda-method}b). \textit{Third}, to validate the learned models, we leverage the cross-validation method in machine learning for a statistical evaluation. The models' classification performance will be assessed by their accuracy on male and female test samples via cross-validation. \textit{Fourth}, we propose a new metric, the Group Specificity Index (GSI), for evaluating the group specificity of the learned models.

Our final contribution involves conducting classification experiments using intrahemispheric connections extracted from resting-state fMRI (rs-fMRI) data of two public neuroimaging repositories, the Human Connectome Project (HCP) \cite{smith2013resting} and the Brain Genomics Superstruct Project (GSP) \cite{holmes2015brain}. The results demonstrate a significant improvement in GSI obtained by GSDA over the baselines.
Further second-order classification reveals consistent sex differences in lateralization across datasets: 1) \textit{about half of the sex-specific lateralized connections are shared between male and female brain functional network, with differences in the strength of lateralization}, 2) \textit{stronger positive inter-lobe interactions are more left-lateralized in the male brain network, while stronger positive intra-lobe interactions are more right-lateralized in the female brain network}.

\section*{Results}\label{sec:res}

\begin{figure*}[t]
    \includegraphics[width=\textwidth]{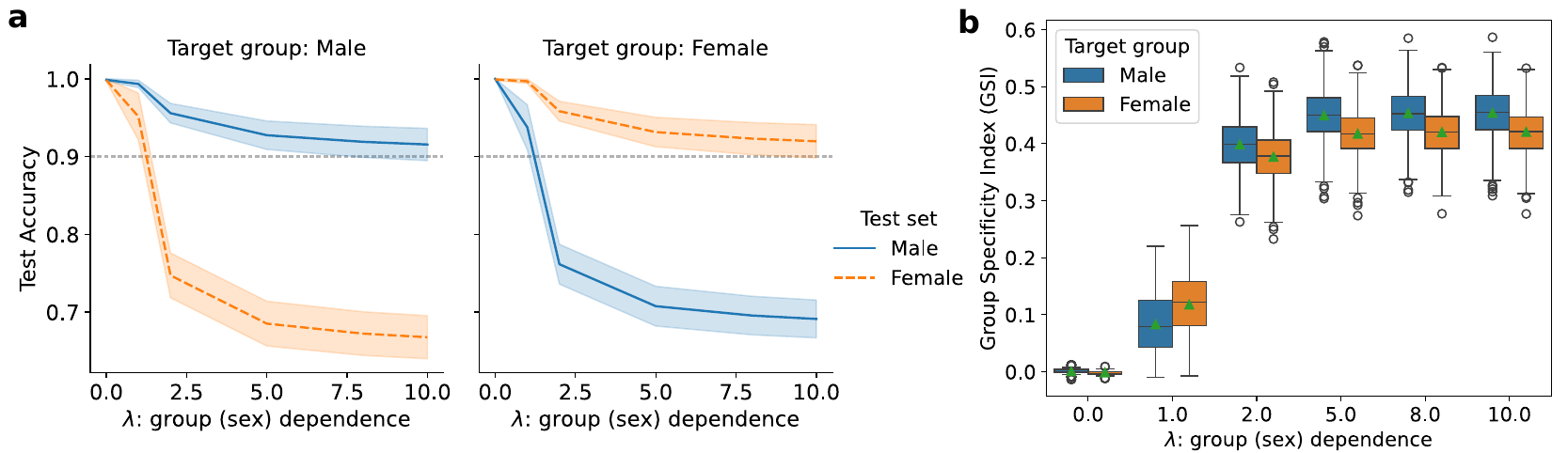}
    \caption{\textbf{Performance of left vs. right brain classification using Group-Specific Discriminant Analysis with the logistic loss (GSDA-Logit) on Human Connectome Project (HCP) \cite{smith2013resting} data} with respect to the hyperparameter $\lambda$, which controls the grouping factor (sex) dependence. A larger $\lambda$ corresponds to a higher dependence. When $\lambda=0$, GSDA-Logit degenerates to a standard logistic regression for the target-group data. There were 1,000 random training-testing partitions for the experiment, where each subject randomly contributed one hemisphere for training and the other for testing, resulting in 50\% brain hemispheres being selected as training samples. \textbf{a}, The test accuracy on male and female sets increasingly diverges with the increase of dependence on sex ($\lambda$). The average test accuracy is represented by solid or dashed lines, with standard deviations shown as error bands, computed across 1,000 random training-testing partitions. \textbf{b}, The Group Specificity Index (GSI) calculated from the test results in Fig. \ref{fig:cls_res}a increases with $\lambda$. The horizontal lines in each box represent the 25th percentile, median, and 75th percentile of the GSI over the 1,000 test sets, respectively, from bottom to top, and the green triangles represent the mean. The average GSI approaches 0 when $\lambda=0$, indicating that the learned logistic regression models without regularization of group dependence are not sex-specific, despite being trained on male (or female) data only.}
\label{fig:cls_res}
\end{figure*}

\subsection*{Diverged test accuracy on male and female sets}

Figure \ref{fig:cls_res} depicts the performance of group-specific discriminant analysis (GSDA) in classifying left vs. right brain hemispheres on the HCP data \cite{smith2013resting}, across a varied range of values for hyperparameter $\lambda$. A larger $\lambda$ indicates a higher grouping factor (sex) dependence. When the target group is male (the left of Fig. \ref{fig:cls_res}a), the labels for the left and right hemispheres of the female training data were masked. Therefore, the training female samples were only involved in the grouping factor dependence regularization. In this scenario, the average accuracy obtained on the male test samples (the blue solid line) stays higher than that on the female test samples (the orange dashed line). The increase of $\lambda$ leads to an increased gap between the test accuracy on target and non-target test sets. In particular, this discrepancy widens significantly within the range $0<\lambda\le 5$ and stabilizes to a 20\% gap for $\lambda > 5$ (Fig. \ref{fig:cls_res}a). These observations remain consistent in results with different cross-validation strategies for the HCP data (\ref{fig:hcp_ext_cls_results}a,c) and the GSP data (\ref{fig:gsp_cls_results}a,c). 

The group specificity of models obtained by GSDA increases with a larger $\lambda$, as reflected by our proposed metric, the Group Specificity Index (GSI), which is presented as a box plot in Fig. \ref{fig:cls_res}b. When $0<\lambda\le 5$, the GSI for both male- and female-specific GSDA models increases with the increase of $\lambda$. When $\lambda\ge 5$, the GSI maintains at around 0.4. Based on both accuracy and GSI results, $\lambda=5$ is an ``elbow'' point in the experiment across different datasets and cross-validation strategies, which can be considered an optimal value for the trade-off between classification accuracy, group specificity, and model complexity (the hyperparameter for $\ell_2$ regularization was fixed to 0.1, so the larger $\lambda$, the lower relative importance of $\ell_2$ regularization). Hence, in the rest of this article, we will use $\lambda = 5$ for GSDA as the main sex-specific model to present the results and findings. 

In contrast, the GSI steadily approaches zero without the grouping factor dependence regularization. At $\lambda=0$, where GSDA degenerates to a standard logistic regression trained only on the target-group hemispheres, the accuracy is nearly 100\% for both male and female test samples (Fig. \ref{fig:cls_res}a, Table \ref{tab:cls}, \ref{fig:hcp_ext_cls_results}a,c, and \ref{fig:gsp_cls_results}a,c). 
This performance is similar to the multivariate control baseline, which uses standard logistic regression trained on mixed male and female hemispheres. From Table \ref{tab:cls}, the control models achieved an accuracy of 99.99 $\pm$ 0.04\% for male and 99.92 $\pm$ 0.13\% for female HCP test samples, and 99.94 $\pm$ 0.07\% for male and 99.99 $\pm$ 0.01\% for female GSP test samples.
Because of the same property and similar performance compared to the standard logistic regression (multivariate control baselines), we will view GSDA with $\lambda=0$ as an additional multivariate baseline. 

\subsection*{GSDA-based models learned distinct weights} 

\begin{figure*}
    \includegraphics[width=\textwidth]{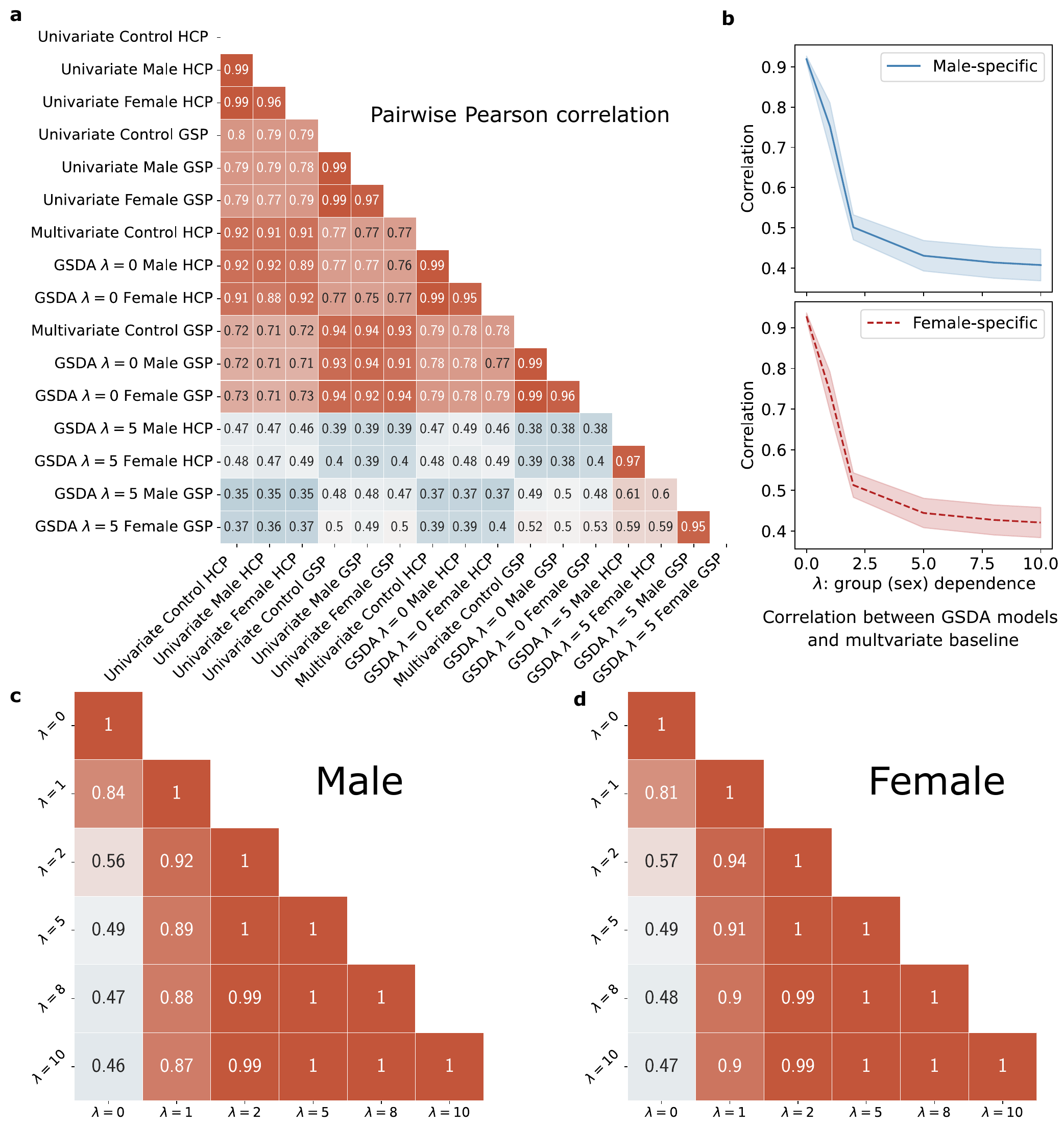}
    \caption{\textbf{Pearson correlation between model weights.} \textbf{a}, Pairwise correlation between weights of 16 models, including multivariate models from Table \ref{tab:cls} and univariate models, labeled along the $x$- and $y$-axis. Two clusters can be observed here: multivariate and univariate baselines versus GSDA with $\lambda=5$. \textbf{b}, Correlation between GSDA and multivariate control models (trained on mixed male and female data) on the HCP data. As $\lambda$ increases, the GSDA models become less correlated with the control models. \textbf{c, d}, Average pairwise correlation for \textbf{c.} male-specific and \textbf{d.} female-specific GSDA models trained on HCP data, with respect to $\lambda$. The weights of sex-specific models remain stable (correlation $\ge 0.99$) for $\lambda \ge 2$.}
	\label{fig:main_corr}
\end{figure*}

Beyond classification performance similarity, the weights of multivariate baselines (control and GSDA with $\lambda=0$) are also highly correlated. As shown in Fig. \ref{fig:main_corr}a, the average Pearson correlation coefficients between multivariate baselines are 0.99 for analyses conducted within either the HCP or GSP datasets. Similarly, in univariate analyses based on the $t$-test of paired left and right connections, the $t$-values of within-group analysis showed a 0.99 correlation with the $t$-values derived from mixed male and female samples (univariate control). Among these multivariate and univariate baselines, the correlation for any arbitrary pair exceeds $0.91$ for within-dataset results and $0.7$ for cross-dataset results. \textit{This high correlation suggests that the lateralization modeled by multivariate or univariate baselines is common to both males and females, regardless of whether the analysis is conducted with exclusively male or female data, or with mixed data.}
This corresponds to the top red triangular cluster in Fig. \ref{fig:main_corr}a.

In contrast, our sex-specific models (with a higher GSI) show lower correlations with the univariate and multivariate baseline models. This corresponds to the blue rectangular cluster at the bottom of Fig. \ref{fig:main_corr}a, where a majority of coefficients fall within the range of 0.35 to 0.5. Increasing the value of $\lambda$ leads to a decreasing correlation between the control and GSDA models ($\lambda>0$), for both results from HCP (Fig. \ref{fig:main_corr}b, and first columns of Fig. \ref{fig:main_corr}c,d) and GSP (\ref{fig:gsp_corr}a,b, and first columns of \ref{fig:gsp_corr}c,d). Moreover, the weights of sex-specific models are stable. As shown in Fig. \ref{fig:main_corr}c,d, the average correlation of any pair for GSDA with $\lambda \ge 2$ is 0.99 or above. 

\begin{table}[t]
    \setlength{\tabcolsep}{4pt}
    \caption{First-order classification (left vs. right brain hemispheres) accuracy on male and female test sets from the HCP \cite{smith2013resting} \& Brain Genomics Superstruct Project (GSP) \cite{holmes2015brain}. Group-specific models (GSDA with $\lambda=5$) are compared with three multivariate baselines: 1) standard logistic regression trained on a mixture of male and female training data, 2) GSDA with $\lambda=0$ (equivalent to standard logistic regression) trained on male data only, and 3) GSDA with $\lambda=0$ trained on female data only. $\lambda=5$ is an optimal value for GSDA on the data as determined by the accuracy and GSI in Fig. \ref{fig:cls_res}. The baselines achieved similar accuracy on both male and female test sets, indicating a lack of group specificity. Conversely, the group-specific models maintained accuracy on the target test set but showed a significant gap with the lower accuracy on the non-target test set.}
    \centering    
    \begin{tabular}{l|ccrccr}
    \toprule
      & \multicolumn{6}{c}{Average test accuracy (\%) and gap (target $-$ non-target)}\\
    
     Classification method (target group) &  HCP male&  HCP female& HCP gap& GSP male& GSP female & GSP gap\\
    \midrule
     Logistic regression (male + female) & 99.99 $\pm$ 0.04 & 99.92 $\pm$ 0.13 & 0.07 & 99.94 $\pm$ 0.07 & 99.99 $\pm$ 0.01 & 0.05 \\
     GSDA ($\lambda=0$, male)& 99.87 $\pm$ 0.16 & 99.85 $\pm$ 0.17 & 0.02 & 99.93 $\pm$ 0.08 & 99.99 $\pm$ 0.01 & 0.06 \\
     GSDA ($\lambda=0$, female)& 99.93 $\pm$ 0.12 & 99.99 $\pm$ 0.04 & 0.06 & 99.97 $\pm$ 0.05 & 99.95 $\pm$ 0.07 & 0.02 \\
     \rowcolor{gray!20}
     GSDA ($\lambda=5$, male) & 92.75 $\pm$ 1.83 & 68.52 $\pm$ 2.88 & 24.23 & 91.85 $\pm$ 1.77 & 71.28 $\pm$ 2.13 & 20.57 \\
     \rowcolor{gray!20}
     GSDA ($\lambda=5$, female)& 70.76 $\pm$ 2.56 & 93.16 $\pm$ 1.89 & 22.40 & 74.70 $\pm$ 2.22 & 92.81 $\pm$ 1.35 & 18.11 \\
    \bottomrule
    \end{tabular}
    \label{tab:cls}
\end{table}

\subsection*{Identifying sex-specific lateralized connections with dual-classification weights}

\begin{figure*}
\centering
 \includegraphics[width=\textwidth]{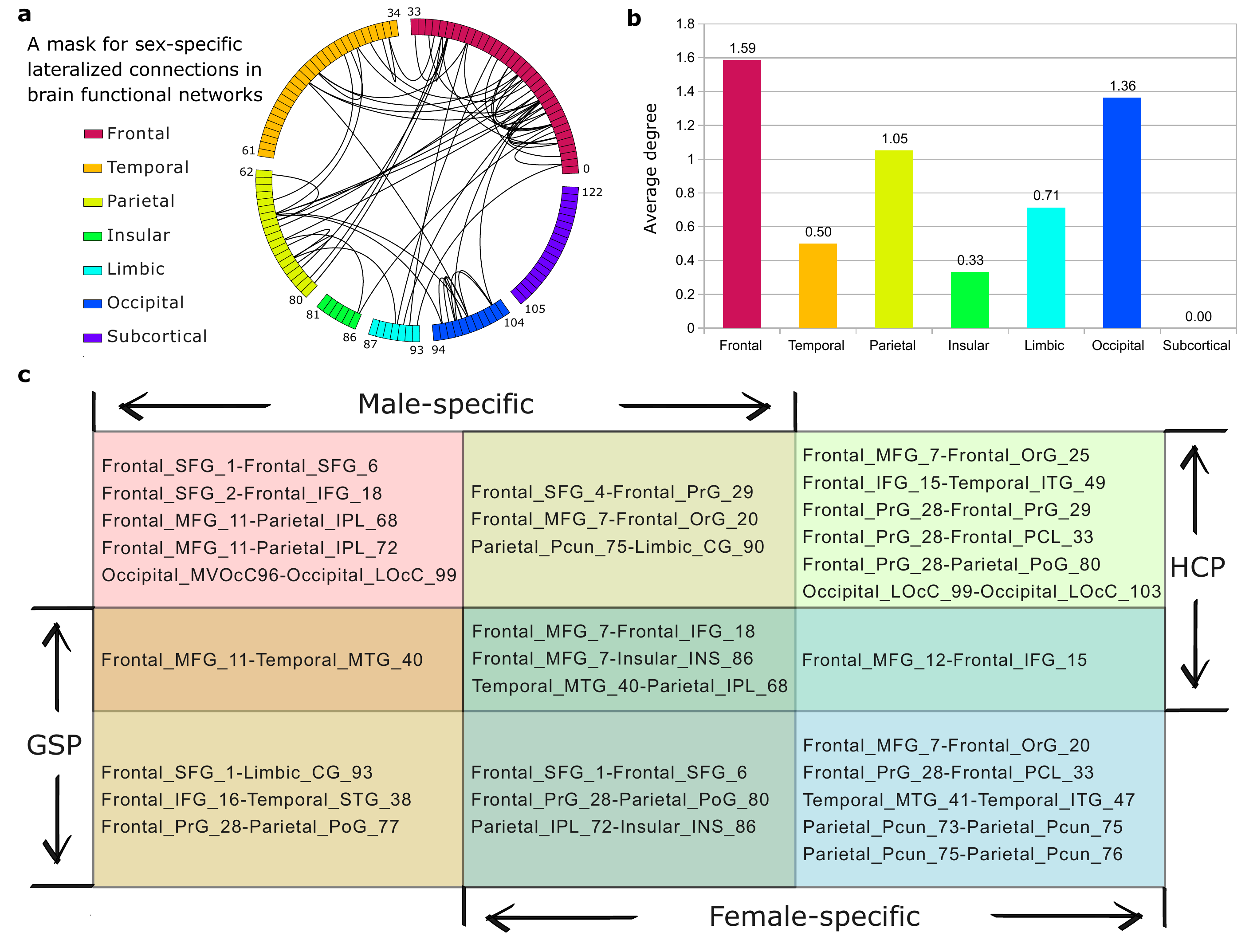}
 \caption{\textbf{Lateralized functional connections identified by sex-specific models (GSDA $\lambda=5$)}. 
 \textbf{a}, A mask for identifying sex-specific lateralized connections is derived in two steps: 1) averaging weights across 1,000 second-order models from different random splits for HCP and GSP data, respectively; 2) identifying overlaps between the top 5\% largest average weights from HCP and those from GSP. The circle represents a brain hemisphere, and each cell on the rim represents a region of interest (ROI) within the half brain. The seven colors indicate seven functional parcellations defined in Brainnetome atlas (BNA) \cite{fan2016human}. The numbers on the rim are the start and end ROI numbers of the lobe in the BNA atlas, where the 123 ROIs are labeled from 0 to 122.
 \textbf{b}, The average degree \cite{diestel2000graph} over ROIs in each of the seven BNA lobes, where the frontal lobe shows the largest degree. 
 \textbf{c}, 30 unique sex-specific lateralized connections learned from HCP and GSP. Each connection is represented in the form of \{ROI\}-\{ROI\}, and each ROI is represented in the form of \{Lobe\}\_\{Gyrus\}\_\{ROI Number\}, where the lobe and gyrus are defined in BNA. The full names of the gyrus are SFG: superior frontal gyrus, MFG: middle frontal gyrus, IFG: inferior frontal gyrus, OrG: orbital gyrus, PrG: precentral gyrus, PCL: paracentral lobule, STG: superior temporal gyrus, MTG: middle temporal gyrus, ITG: inferior temporal gyrus, IPL: inferior parietal lobe, Pcun: precuneus, PoG, postcentral gyrus, INS: insular gyrus, CG: cingulate gyrus, MVOcC: medioVentral occipital cortex, LOcC: lateral occipital cortex. 22 out of 30 connections are associated with the frontal. These connections are either shared between males and females (middle column) or ``exclusive'' to one group (left and right column). For each sex-specific model, e.g. male-specific model on HCP, about half of the identified connections are shared. Nine connections are shared in total and their sex differences are in the strength of lateralization (Fig. \ref{fig:edges}a-d). 21 connections are ``exclusive'' and their sex differences are in the patterns of inter-/intra-lobe interactions (Fig. \ref{fig:edges}e-f, \ref{fig:rader}).
 }
 \label{fig:chord}
\end{figure*}

\begin{figure*}[t]
\centering
 \includegraphics[width=\textwidth]{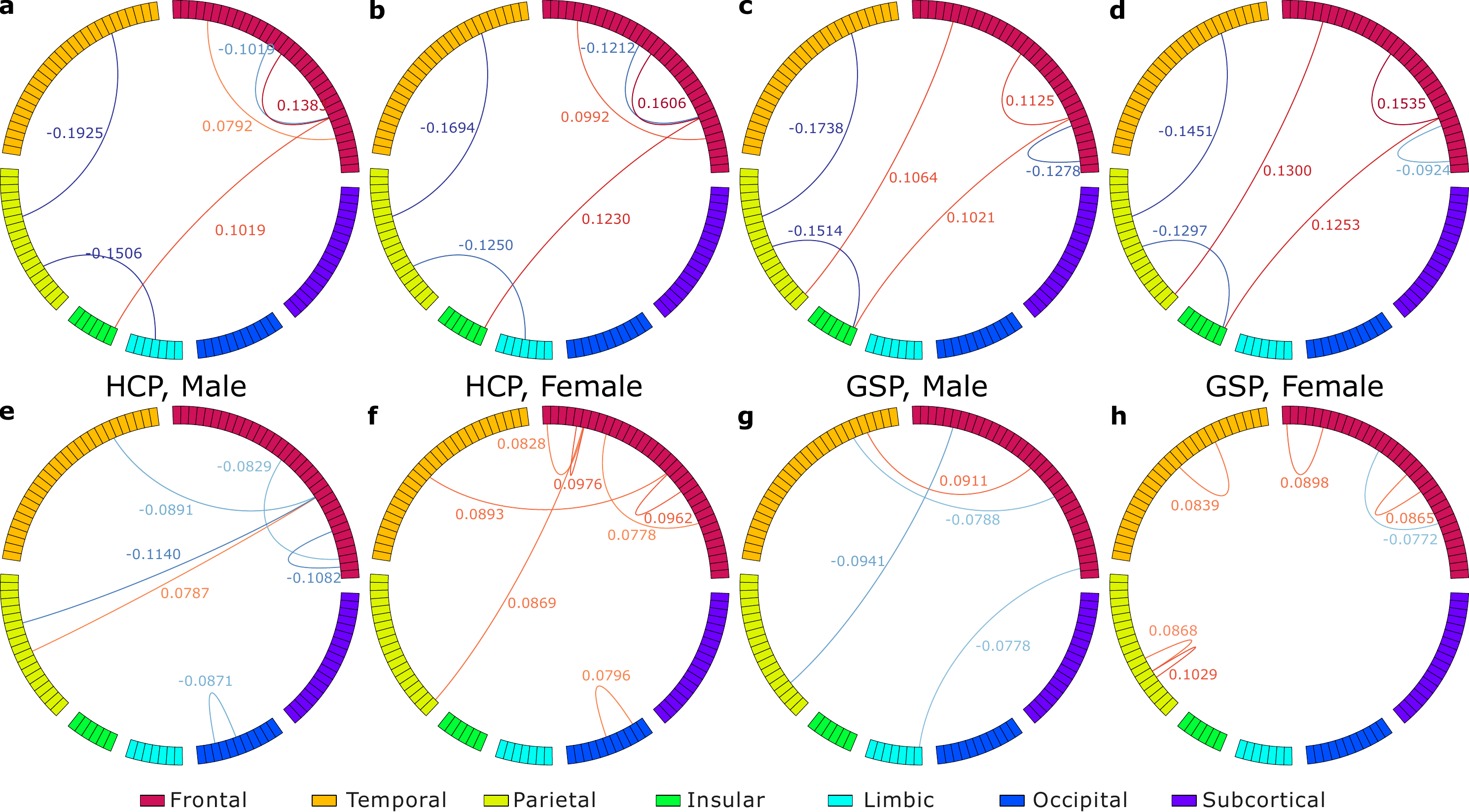}
 \caption{\textbf{Sex-specific lateralized connections with first-order model weights (GSDA $\lambda=5$)}. The connections were identified by applying the mask in Fig. \ref{fig:chord}a to the top 5\% weights (by magnitude) from four models specific to males of HCP (\textbf{a} and \textbf{e}), females of HCP (\textbf{b} and \textbf{f}), males of GSP (\textbf{c} and \textbf{g}), and females of GSP (\textbf{d} and \textbf{h}). Each of the four models was obtained by averaging the corresponding 1,000 first-order models. The sex-specific lateralized connections consist of shared connections between male and female models (\textbf{a}-\textbf{d}), and the group (sex) ``exclusive'' connections (\textbf{e}-\textbf{h}).
    The weights of these shared connections show consistent sex differences.
    In female-specific models, the weights for connections involving the frontal lobe tend to be larger than those in male-specific models, especially for positive weights. Conversely, in male-specific models, the weights for connections to other lobes are generally larger than those in female-specific models.
    The ``exclusive'' connections in male-specific models are mostly inter-lobe and negative, whereas in female-specific models, they are mostly intra-lobe and positive. Statistics about these connections can be found in Fig. \ref{fig:rader}.
 }
 \label{fig:edges}
\end{figure*}

\begin{figure*}[t]
\centering
    \includegraphics[width=0.85\textwidth]{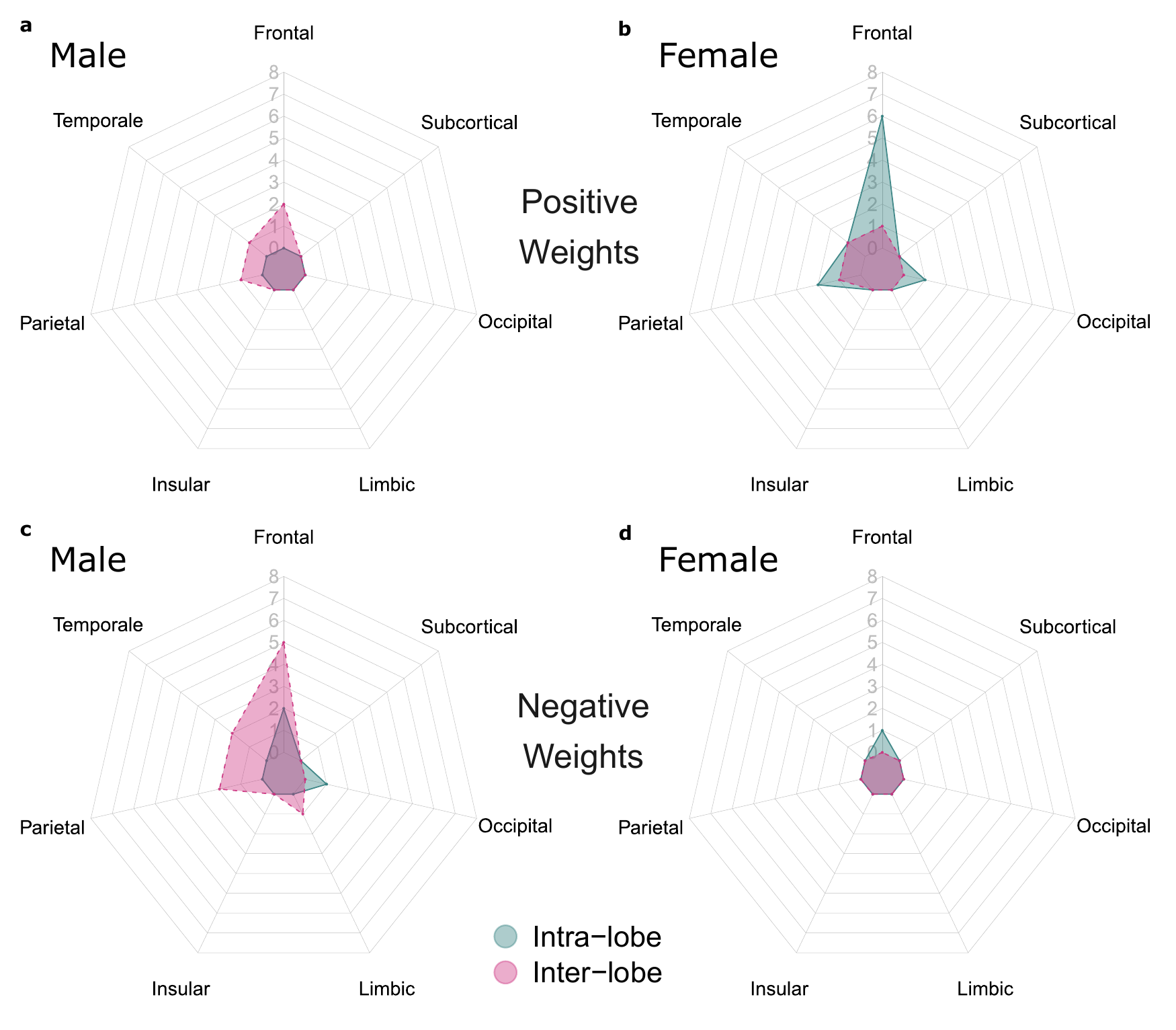}
    \caption{\textbf{Count of the group ``exclusive'' lateralized connections for HCP and GSP (Fig. \ref{fig:edges}e-h), categorized by associated lobes, inter- or intra-lobe, and signs of the first-order weights.} The connections were identified by 
    \textbf{a}, the male-specific models with positive first-order weights; 
    \textbf{b}, the female-specific models with positive first-order weights;
    \textbf{c}, the male-specific models with negative first-order weights;  
    \textbf{d}, the female-specific models with negative first-order weights.
    In male brain networks, 7 out of 10 ``exclusive'' connection counts are inter-lobe, with 71.4\% of the inter-lobe connections having negative first-order weights. In female brain networks, 11 out of 13 ``exclusive'' connection counts are intra-lobe, with 91.7\% of the intra-lobe connections having positive first-order weights.}
 \label{fig:rader}
\end{figure*}

To identify lateralized connections related to sex differences among the 7,503 intrahemispheric connections, we performed a second-order classification. This involved training standard logistic regression models to distinguish between male- and female-specific models learned from the first-order classification, using 80\% of the first-order models for training and 20\% for testing.
The test accuracy for second-order classification consistently achieved nearly 100\% over 1,000 random splits. This indicates that the sex differences in the first-order model weights are generalizable. 

Based on the weights from these second-order classification models, we derived a mask that characterizes sex differences in the lateralized connections. We first averaged the weights across 1,000 second-order classification models from different random splits for the HCP and GSP datasets, respectively.
Then, we identified the overlap between the top 5\% of the largest average weights (by magnitude) from HCP and those from GSP. The resulting map is represented by the figure of chords in Fig. \ref{fig:chord}a. The threshold of 5\% was chosen because the second-order logistic regression classifiers were trained with $\ell_2$ regularization, which can be interpreted as a Gaussian prior (normal distribution) on model weights, with 5\% being a commonly used statistical significance level for a Gaussian distribution. By calculating the average degree \cite{diestel2000graph} (Fig. \ref{fig:chord}b) of connections for each lobe within this mask, we can learn that sex differences in first-order weights are associated with frontal, parietal, and occipital lobes, where the average degrees exceed 1. The frontal lobe shows the largest average degree, indicating significant sex differences.

We then applied this mask to the top 5\% weights of four first-order classification models: HCP male-specific, HCP female-specific, GSP male-specific, and GSP female-specific. The obtained lateralized connections with sex differences are shown in Fig. \ref{fig:chord}c, Fig. \ref{fig:edges}a-h, and \ref{fig:chord-ext}a-d. The weights of these four models were obtained by taking the average of the corresponding 1,000 models learned from different random splits. In total, 47 lateralized connections with repetition were identified, of which 30 connections are unique. Among these 47 connections, the middle frontal gyrus (MFG) was the most frequently involved region, suggesting it may serve as a hub. Specifically, 17 out of the 47 connections were associated with the MFG in both male and female samples across both datasets.

\subsection*{Sex-specific lateralization: shared and ``exclusive'' connections}
For each of the four sex-specific models, half of the identified lateralized connections are shared between male and female brain networks (Fig. \ref{fig:chord}c) on average: on HCP, 6 out of 12 for the male-specific and 6 out of 13 for the female-specific model; on GSP, 6 out of 10 for the male-specific and 6 out of 12 for the female-specific model. Among the 30 unique sex-specific lateralized connections identified across datasets, 9 (nearly one-third) are shared between males and females. To illustrate these findings, we have separated the shared and ``exclusive'' connections, as depicted in Fig. \ref{fig:edges}a-h and Fig. \ref{fig:edges}e-h, respectively. 

For the shared lateralized connections (Fig. \ref{fig:edges}a,b for HCP and Fig. \ref{fig:edges}c,d for GSP), we observed sex differences in the magnitude of first-order weights, i.e., strength of lateralization. Specifically, for the female-specific models, the magnitudes of first-order weights corresponding to the connections associated with the frontal lobe are generally larger compared to those for male-specific models, particularly those of the positive weights. In male-specific models, the magnitudes of first-order weights for connections related to other lobes are larger than those in female-specific models.

For the ``exclusive'' connections in the four models (Fig. \ref{fig:edges}e-h), male-specific models contain more inter-lobe lateralized connections (Fig. \ref{fig:rader}a), with more than 70\% of corresponding weights being negative, as shown by the blue chords in Fig. \ref{fig:edges}e,g. Female-specific models, on the other hand, contain more intra-lobe lateralized connections (Fig. \ref{fig:rader}b), with more than 90\% of weights being positive, indicated by the red chords in Fig. \ref{fig:edges}f,h. Notably, these patterns of inter- and intra-lobe lateralization for males and females are consistent across joint or separate analyses of both HCP and GSP data (Fig. \ref{fig:rader-ext}), demonstrating the stability and reliability of these findings. 

\section*{Discussion}

\subsection*{Cross-validation challenges conventional statistical approach for investigating lateralization}

Traditional neuroscience studies commonly assume that results from within-group analyses are specific to the group being studied \cite{schwarz2011sex, kret2011men, zullo2020gene, jacobsen2007gender, cui2018individualized}. This methodology of exclusively using male or female data to explore sex-specific characteristics is intuitively logical. However, our experimental results obtained via cross-validation challenge this assumption. For example, as depicted in Fig. \ref{fig:cls_res}a, when standard logistic regression models are exclusively trained on data from one target group (male or female), the performance on test sets for both the target and non-target groups is nearly identical. Hence, the performance of these baseline models is not sensitive to sex-based sampling. Based on the definition of generalization in statistical learning theory \cite{vapnik1999nature}, the similarity in generalization errors suggests that these models are general to both males and females, not sex-specific. This finding implies that statistical methods can learn common patterns even when trained on data from a specific group, contradicting the conventional assumption in group-specific analysis. This conclusion holds at least in our study on left vs. right brain classification using the HCP and GSP datasets.

The performance of our sex-specific models (GSDA with $\lambda>0$) is sensitive to sex-based sampling. The classification results (Fig. \ref{fig:cls_res}a, \ref{fig:hcp_ext_cls_results}a,c, and \ref{fig:gsp_cls_results}a,c) reveal that the generalization error for the target-group test sets is significantly lower than that for the non-target-group test sets. This indicates a stronger specificity to sex compared to the multivariate baseline models, as reflected by our group-specificity index (GSI) results.
The differences observed in test performance highlight the importance of cross-validation in validating the group specificity of statistical analysis results. 

While univariate analysis results are not directly applicable to unseen samples for testing, the strong correlation between univariate and multivariate baselines offers valuable insights for pattern identification. 
For example, the correlation of the within-group univariate $t$-test results with univariate control models (mixed) and multivariate baselines exceeds 0.99 and 0.91, respectively. This suggests that the outcomes of our within-group univariate analyses are likely common to both males and females. Consequently, previous conclusions from such within-group analyses should be revisited and validated. Moreover, although multivariate methods are theoretically superior in capturing interactions between features compared to univariate methods, the observed similarity between multivariate and univariate results suggests that multivariate methods might not always identify patterns distinct from those found by univariate methods.

\subsection*{Regions and connections identified across datasets for sex-specific lateralization}

The mask resulting from the second-order classification revealed sex differences in connections across lobes, including the frontal, temporal, parietal, insular, limbic, and occipital lobes, where functional differences between males and females were observed in previous studies \cite{agcaoglu2015lateralization, smith2014characterizing, reber2017sex, zhang2018functional, sen2020predicting,kong2019spatial}.
Among the sex-specific lateralized connections, MTG-IPL, MFG-IFG, and MFG-INS are shared in both male- and female-specific models across the two datasets. 

\textit{From the perspective of gyrus}, which engages in various cognitive functions, the lateralized regions include the MTG\cite{onitsuka2004middle} (sound recognition and language processing), MFG \cite{koyama2017differential} (literacy and numeracy), IPL \cite{lynch1980functional,hyva1981regional} (spatial attention, multimodal sensory integration, and oculomotor control), IFG\cite{greenlee2007functional} (speech and language processing), and INS \cite{kortz2021insular} (various sensorimotor processing and risk-reward behavior). 
These regions show lateralization and sex differences in certain functions including speech processing, language, and spatial attention \cite{hirnstein2019cognitive, agcaoglu2015lateralization, smith2014characterizing}. The MFG, a hub region in this study, is a core component of the multiple demand system \cite{duncan2010multiple}, and presents hemispheric specialization, with the left MFG primarily supporting literacy development, while the right MFG is vital for numeracy \cite{koyama2017differential}. 
The MTG showed lateralization in activated volumes for both males and females during language tasks, while the lateralization of IFG was only observed in males \cite{kansaku2000sex}. Our study reports different weights of connections related to these two regions. This suggests that the lateralization of a region's external connections can reflect the lateralization of its functional activation. The ``activity flow'' theory in neuroscience has linked the connections and functional activation \cite{cole2016activity}, proposing that the seed-based connection-weighted sums of the activation of other regions can predict functional activation of the seed region. Our results suggest \textit{a correlation between the lateralization of functional connectivity and activation}, although further quantitative analysis is required to investigate the specifics of this relationship. 

\textit{From the perspective of connections}, 
the MTG-IPL connection is associated with picture naming and displays notable plasticity \cite{van2019resting}. Laws (2004) \cite{laws2004sex} assessed sex differences in picture naming speed, and Ala-Salomäki \textit{et al.} \cite{ala2021picture} found that picture naming shows reliable left-lateralized evoked activation. These studies indicate the potential of sex and hemispheric differences for MTG-IPL connection and picture-naming cognition. Our GSDA framework successfully captured these effects, suggesting that the underlying mechanism for picture naming lateralization could be the lateralization of related connections, such as the MTG-IPL connection. As for the MFG-IFG connection, proficiency in artificial grammar rules was found to be positively linked to the functional connectivity between the left IFG and left MFG \cite{kepinska2018connectivity, kansaku2000sex}, and this function also identified sex differences \cite{lotz2011rapid}. In consistency with these studies, our framework has captured both the lateralization and sex effects on the MFG-IFG connection. 
In the case of the MFG-INS connection, associated with mild traumatic brain injury (mTBI)\cite{li2022causal}, our study reveals sex-specific hemispheric differences, suggesting that personalized intervention and diagnostic approaches considering brain hemispheric and sex differences may be necessary for more effective mTBI treatment.

\subsection*{Sex differences in lateralization: strength and inter-/intra-lobe interaction patterns}
\textit{Sex differences in strengths of shared lateralized connections}: As reported in the results section, the first-order weights corresponding to these shared connections show consistent sex differences (Fig. \ref{fig:chord}c-f). In our labeling strategy, ``left'' was labeled as 0 and ``right'' as 1. Therefore, a positive first-order weight indicates that stronger positive interactions (FC value approaching 1) between two ROIs suggest a higher probability of right lateralization. Conversely, a more negative interaction (FC value approaching -1) indicates that stronger negative interactions between two ROIs suggest a higher probability of left lateralization. The opposite interpretation applies to negative first-order weights. Therefore, we can interpret the sex differences in first-order weights for the shared connections as follows: \textit{positive interactions involving the frontal lobes are more right-lateralized in females than in males, whereas positive interactions involving the temporal, parietal, insular, and limbic lobes are more left-lateralized in males than in females}. This observation of difference in strength of lateralization aligns with the findings on the lateralization patterns of right- and left-handed individuals \cite{vingerhoets2012cerebral}, and supports neuroscience discoveries of shared functional network mechanisms across males and females \cite{linnman2012sex, cui2018individualized}. 

\textit{Sex differences in inter-/intra-lobe interaction patterns are identified by the ``exclusive'' lateralized connections,} particularly within the frontal lobe (Fig. \ref{fig:chord}g-j), a key region for language processing \cite{tuckute2022frontal}. Using the same approach as above for interpreting first-order weights, we can summarize that \textit{males tend to have a stronger left lateralization in positive inter-lobe interactions, while females tend to have a stronger right lateralization in positive intra-lobe interactions}. This divergence may result from the evolutionary pressure for lateralization, which optimizes functional organization and reduces redundancy among brain regions \cite{corballis2019evolution, vallortigara2005survival}. Inter-lobe connections, characterized by long-range wiring, are metabolically costly \cite{laughlin2003communication, lennie2003cost, levy1996energy, ringo1994time}, while the shorter-range intra-lobe connections are more energy-efficient. In females, these intra-frontal connections may enhance language abilities. Conversely, male inter-lobe connections may be driven by the need to engage more extensive functional areas for complex visuospatial tasks. This divergence can be a factor in sex and lateralized differences in cognitive abilities, with males typically outperforming in rightward visuospatial tasks, and females in leftward verbal tasks \cite{levy1972lateral,levy1978lateral}. 

\subsection*{Potential of GSDA for more general applications}

Our study focused on sex as a grouping factor and employed brain hemisphere labels to identify sex-specific lateralized patterns for human brain functional networks. The results demonstrate efficacy and stability in identifying and validating sex differences in lateralization. Importantly, the scope of this general predictive framework extends beyond its current application. For instance, employing different grouping factors, such as handedness or disease vs. control, can advance the analysis of distinct characteristics. Moreover, the GSDA algorithm can be extended to grouping factor combinations, such as sex and handedness. While this work focused on classification and a discrete group factor, our method can be adapted for regression tasks (e.g. predicting behavioral/cognitive scores) and continuous grouping factors (e.g. age or IQ score).

\bibliography{main_ref}

\newpage

\nocitesupp{*}

\section*{Methods}

\subsection*{Dual classification with group-specific discriminant analysis (GSDA)}
We propose a dual classification framework with two primary objectives: learning group-specific models and identifying group-specific discriminant weights. For the first objective, specifically the classification of left vs. right brain hemispheres, we train a linear classifier on the training data and then validate its performance on the test data. We refer to this process as the \textbf{first-order classification}. The weights derived from the model are called the \textbf{first-order weights}. Then we perform a second round of classification to identify the weights that show significant differences between group-specific models. Here, we train a linear classifier to differentiate between male- and female-specific models. This stage is called the \textbf{second-order classification}, and the associated weights are referred to as the \textbf{second-order weights}. This process is illustrated in stages \ding{176} and \ding{177} of Fig. \ref{fig:gsda-method}a.

The first-order classification builds a (group-specific) prediction function. This function predicts whether an unseen brain hemisphere is left or right, based on a feature vector. These vectors represent the left or right human brain hemispheres and are extracted from the training neuroimaging data. The resulting prediction accuracy serves as a quantitative measure of the extent to which the learned lateralization patterns are generalized among the brain networks within the test set. The learned model weights can be interpreted as indicators of the significance or extent of differences between the corresponding connections of the left and right brain hemispheres.

The second-order classification is designed to identify weights that show significant differences between the male- and female-specific first-order models. In this stage, a linear classification model is trained on the first-order model weights to predict whether an unseen model is male- or female-specific. The features with larger weights in the second-order classification are considered to represent the stronger sex differences.
We propose a group-specific discriminant analysis algorithm to learn group-specific models for the first-order classification. 

\subsubsection*{Problem formulation of GSDA}
Let $(\mathbf{x}_i, y_i, \mathbf{g}_i)$ represent the $i$th sample, where $\mathbf{x}_i \in \mathcal{X} \subseteq \mathbb{R}^p$ denotes an input data vector, $y_i \in \mathcal{Y}$ denotes an output variable (label), and $\mathbf{g}_i \in \mathcal{G} \subseteq \mathbb{R}^q$ represents a covariate vector for the grouping factor(s). Here, $i \in [1, m]$, with $m$ being the total number of samples. $\mathcal{X}$, $\mathcal{Y}$ and $\mathcal{G}$ are the feature spaces of the input data, output label, and grouping factor, respectively, with $p$ and $q$ as the corresponding feature dimensions for the input data $\mathbf{x}_i$ and grouping factor $\mathbf{g}_i$. In the context of this article, $\mathbf{x}_i$ is a feature vector that represents a brain hemisphere, $y_i$ indicates whether $\mathbf{x}_i$ is the left or right hemisphere, and $\mathbf{g}_i$ is a binary (zero and one) indicator representing whether $\mathbf{x}_i$ is from a male or female subject (e.g. $g_i = 0$ for male and $g_i = 1$ for female).
Assuming $\mathbf{x}_0 = 1$, considering $\mathbf{w}_0$ as the bias term, and denoting $\mathbf{w} \in \mathbb{R}^{p+1}$ as the vector of weights (coefficients) to be learned, with the target group represented as subscript $_\mathrm{t}$, we formulate the objective of learning group-specific models as follows:
\begin{equation}
	\argmax_{\mathbf{w}} \frac{1}{m_\mathrm{t}}\sum_{i=1}^{m_\mathrm{t}} \mathbb{P}(y_i|\mathbf{x}_i, \mathbf{w}) +  \frac{\lambda}{m}\sum_{j=1}^{m}  \mathbb{P}(\mathbf{g}_j|\mathbf{x}_j,\mathbf{w}),
	\label{eq:obj}
\end{equation}
where $m_\mathrm{t}$ denotes the number of training samples from the target group, and $\lambda \ge 0$ is the hyperparameter that quantifies the importance of grouping factor(s) dependence.
Based on Eq. (\ref{eq:obj}), we formulate a general group-specific discriminant analysis (GSDA) framework as:
\begin{equation}\label{eq:gsda-framework}
    \argmin_{\mathbf{w}} L(\mathbf{X}_{\mathrm{t}}^\top\mathbf{w},\mathbf{y}_{\mathrm{t}}) + \alpha \|\mathbf{w}\|_K^2 - \lambda \underbrace{\rho(\mathbf{X}^\top\mathbf{w}, \mathbf{G})}_{\text{Group dependence}},
\end{equation}
where $L(\cdot, \cdot)$ denotes a classification or regression loss function, such as least square, logistic, or hinge, $ \alpha \ge 0 $ is the hyperparameter used for weight regularization, $\|\cdot\|_K^2$ denotes either an $\ell_1$ or $\ell_2$ regularization, with $K=1$ or $2$ respectively, $\mathbf{X}_{\mathrm{t}}$ denotes the target group's training samples, $\mathbf{X}$ denotes all training samples that consist of both target and non-target group samples, and $ \rho(\cdot, \cdot) $ is a statistical dependence measure. In this work, we employed \textit{Hilbert-Schmidt Independence Criterion (HSIC)} \citesupp{gretton2005measuring}. Given two sets $ \mathbf{X}=\{\mathbf{x}_1, \mathbf{x}_2, \dots, \mathbf{x}_m\} $ and $ \mathbf{Y}=\{\mathbf{y}_1, \mathbf{y}_2,\dots, \mathbf{y}_m\} $, both with size $m$, HSIC compute the statistical dependence between tests whether $\mathbf{X}$ and $\mathbf{Y}$ via 
\begin{equation}
	\rho_h(\mathbf{X, Y}) = \frac{1}{(m-1)^{2}}\text{tr}(\mathbf{KHLH}),
	\label{eq:hsic}
\end{equation}
where $ \mathbf{K, H, L} \in \mathbb{R}^{m\times m} $, $ \mathbf{K}_{i,j}:= k_x(\mathbf{x}_i, \mathbf{x}_j) $, $ \mathbf{L}_{i,j}:= k_y(\mathbf{y}_i, \mathbf{y}_j) $, $k_x(\cdot, \cdot)$ and $k_y(\cdot, \cdot)$ are two kernel functions, such as linear, polynomial, or radial basis function (RBF), $ \mathbf{H} = \mathbf{I} -\frac{1}{m}\mathbf{11^\top} $ is the centering matrix, $\mathbf{I}$ is an identity matrix, and $\text{tr}\left(\cdot\right)$ is the trace function. HSIC $\rho(\mathbf{X, Y})\ge 0$, and it is zero if and only if the two sets of variables $ \mathbf{X}$ and $ \mathbf{Y} $ are independent, i.e., $\mathbb{P}(\mathbf{x, y}) = \mathbb{P}(\mathbf{x})\mathbb{P}(\mathbf{y})$. A higher HSIC value suggests stronger statistical dependence.

\subsubsection*{GSDA with logistic loss and maximum likelihood estimation}

To maximize the likelihood of the target-group labels and the grouping factor(s) dependence as specified in Eq. (\ref{eq:obj}), we adopt maximum likelihood estimation for optimizing the model weights $\mathbf{w}$. Here, we develop a novel algorithm, Group-Specific Discriminant Analysis with logistic loss (GSDA-Logit), as a variant of logistic regression for group-dependent learning. The overall likelihood to be maximized is as follows:
\begin{align}
\begin{aligned}\label{eq:GSDA-Logit_max_likelihood}
    \mathbb{P}(\mathbf{y}_{\mathrm{t}}|\mathbf{X}_{\mathrm{t}},\mathbf{w})\mathbb{P}(\mathbf{w})\mathbb{P}(\mathbf{G}|\mathbf{X},\mathbf{w}) = \big(\prod_{i=1}^{m_{\mathrm{t}}}S(\mathbf{w^\top x}_i)^{y_i}(1-S(\mathbf{w^\top x}_i))^{(1 - y_i)} \big) \frac{1}{\sqrt{2\pi \sigma}}e\big(-\frac{\mathbf{w^\top w}}{2\sigma^2}\big) S(\rho_h(\mathbf{w^\top X}, \mathbf{G})),
    \end{aligned}
\end{align}
where $\mathbb{P}(\mathbf{y}_{\mathrm{t}}|\mathbf{X}_{\mathrm{t}}, \mathbf{w})$ is the likelihood of target labels $\mathbf{y}_{\mathrm{t}}$ given the model and target-group data $\mathbf{X}_{\mathrm{t}}$, $\mathbb{P}(\mathbf{w})$ is the prior probability of weights in $\mathbf{w}$, $\mathbb{P}(\mathbf{G}|\mathbf{X}, \mathbf{w})$ is the likelihood of grouping factor dependence, $S(\cdot)$ denotes the logistic (or sigmoid) function, and $ \mathbb{P}(\mathbf{w}) $ can be interpreted as the $ \ell_2 $ regularization for $ \mathbf{w} $, assuming that the weights in $ \mathbf{w} $ follow a normal distribution with a mean of zero and a standard deviation of $ \sigma $. Given that $\mathbf{w^\top X}$ produces a row vector, Equation (\ref{eq:hsic}) can be reformulated in the form of simplified HSIC \citesupp{zhou2020side}:
\begin{align}
    \begin{aligned}
        \rho_{sh}(\mathbf{w^\top X}, \mathbf{G})& = \text{tr}((\mathbf{w^\top X})^\top(\mathbf{w^\top X})\mathbf{HL}\mathbf{H}) \\
        &= \mathbf{w}^\top\mathbf{XHLHX}^\top\mathbf{w},
    \end{aligned}
    \label{eq:HSIC_simple}
\end{align}
where $ \mathbf{L} = \mathbf{G^\top G} $. By replacing $\rho_h(\mathbf{w^\top X}, \mathbf{G})$ with the simplified HSIC $\rho_{sh}(\mathbf{w^\top X}, \mathbf{G})$, the likelihood can be rewritten as
\begin{align}
	\begin{aligned}\label{eq:GSDA-Logit_max_likelihood_hsic}
		\mathbb{P}(\mathbf{y}_{\mathrm{t}}|\mathbf{X}_{\mathrm{t}},\mathbf{w})\mathbb{P}(\mathbf{w})\mathbb{P}(\mathbf{G}|\mathbf{X},\mathbf{w}) = \big(\prod_{i=1}^{m_{\mathrm{t}}}S(\mathbf{w^\top x}_i)^{y_i}(1-S(\mathbf{w^\top x}_i))^{(1 - y_i)} \big) \frac{1}{\sqrt{2\pi \sigma}}e\big(-\frac{\mathbf{w^\top w}}{2\sigma^2}\big) S(\mathbf{w^\top XHLHX^\top w}).
	\end{aligned}
\end{align}
The likelihood in Eq. (\ref{eq:GSDA-Logit_max_likelihood_hsic}) can be maximized using the same optimization steps for a standard logistic regression, i.e., computing the gradient of the negative log-likelihood. Let $ \alpha = \frac{1}{\sigma^2}$ and $ \lambda $ denote the two hyperparameters that control the importance of the $ \ell_2 $ regularization and grouping factor dependence regularization, respectively. Let $J(\mathbf{w})$ denote the negative logarithm of the likelihood. Taking the gradient of $J(\mathbf{w})$ with respect to $\mathbf{w}$, we obtain
\begin{align}
    \begin{aligned}\label{eq:GSDA-Logit_gradient}
	\nabla J(\mathbf{w}) = \mathbf{X}_{\mathrm{t}}(S(\mathbf{X}^\top_{\mathrm{t}}\mathbf{w}) - \mathbf{y}_{\mathrm{t}}) + \alpha \mathbf{w} + \lambda(S(\mathbf{w^\top XHLHX^\top w})-1)\mathbf{XHLHX^\top w}.
\end{aligned}
\end{align}
Finally, $ \mathbf{w} $ can be optimized iteratively via
\begin{equation}
    \mathbf{w}^{k+1} = \mathbf{w}^k - \eta \nabla J(\mathbf{w}^k), 
\end{equation}
where $ k $ denotes the $ k $th iteration, $ \eta $ is the learning rate (step size). Algorithm \ref{alg:GSDA-Logit} is the pseudocode for GSDA-Logit. In addition to standard gradient descent optimization, we have implemented the LBFGS optimization algorithm \citesupp{liu1989limited} for faster coefficient estimation.

\begin{algorithm}[tb]
    \caption{Group-Specific Discriminant Analysis with logistic loss (GSDA-Logit)}
    \label{alg:GSDA-Logit}
    \textbf{Input}: Input data matrix $ \mathbf{X}\in \mathbb{R}^{p \times m}  $, target-group label vector $ \mathbf{y}_{\mathrm{t}}\in \mathbb{R}^{m_{\mathrm{t}}}$, grouping factor(s), and indices of samples from the target group (optional, if not given, first $m_{\mathrm{t}} < m$ samples are assumed to be the labeled target samples).\\
    \textbf{hyperparameters}: $ \alpha $ for $ \ell_2 $ regularization, $ \lambda $ for group dependence (HSIC) regularization, and $ \eta $ for learning rate. \\
    \textbf{Output}: Coefficient vector $ \mathbf{w} \in \mathbb{R}^{p+1} $. 
    \begin{algorithmic}[1] 
        \State Encode the grouping factor(s) into a matrix $ \mathbf{G}\in \mathbb{R}^{q\times m}$ ($q=1$ for a binary grouping factor) using one-hot encoding. Then construct the kernel matrix $\mathbf{L} \in \mathbb{R}^{m \times m} = \mathbf{G}^\top \mathbf{G}$ and the centering matrix $\mathbf{H}\in \mathbb{R}^{m \times m}$;
        \State Add a row of $ 1 $s to $ \mathbf{X} $;
        \State Randomly initialize $\mathbf{w}^{k} $ ($k=0$);
        \While{Not converge}
        \State Compute gradient $ \nabla J(\mathbf{w}) $ by Eq. (\ref{eq:GSDA-Logit_gradient});
        \State Update $ \mathbf{w}^{k+1} = \mathbf{w}^k - \eta \nabla J(\mathbf{w}^k) $;
        \EndWhile
        \State \textbf{return} GSDA-Logit coefficient vector $ \mathbf{w} $.
    \end{algorithmic}		
\end{algorithm}

\subsubsection*{Evaluation: group specificity index (GSI)}
To measure the group specificity, we set the following criteria for the metric
\begin{itemize}
    \item The value of this metric lies within $[0, 1]$.
    \item The value of this metric is 0 if the test accuracy for the target and non-target groups are identical.
    \item When the test accuracy of the target and non-target groups differ, the value of this metric should be proportional to 1) the absolute accuracy for the target group, and 2) the closeness of accuracy for the non-target group to the random chance.
    \item Greater relative accuracy divergence between target and non-target groups will result in a higher value of this metric.
\end{itemize}
To satisfy the above conditions, we propose a Group Specificity Index (GSI) for binary classification problems as follows:
\begin{equation}\label{eq:gsi}
    \mathrm{GSI} = 2\mathrm{BAT}(\mathrm{BAT} - 0.5 - |\mathrm{BANT} -0.5| ),
\end{equation}
where $\mathrm{BAT} \in [0.5, 1]$ represents the balanced accuracy of the target-group data, and $\mathrm{BANT} \in [0, 1]$ denotes the balanced accuracy of the non-target-group data. Balanced accuracy is chosen to mitigate the impact of imbalanced samples. It is defined as $\mathrm{BA} = \frac{\mathrm{TPR} + \mathrm{TNR}}{2}$, where the true positive rate $\mathrm{TPR} = \frac{\text{number of true positives}}{\text{number of total positives}}$ and the true negative rate $\mathrm{TNR} = \frac{\text{number of true negatives}}{\text{number of total negatives}}$. In the left vs. right brain hemisphere classification problem, the numbers of left and right training examples are equal, making balanced accuracy equivalent to accuracy. The expression $|\mathrm{BANT} - 0.5|$ measures how close the accuracy of the non-target group is to random chance (0.5), and $\mathrm{BAT} - 0.5 - |\mathrm{BANT} - 0.5|$ quantifies the relative accuracy divergence between the target and non-target groups. Since we are interested in generalized lateralization patterns for the target group, models that perform worse than random chance on target test sets are not considered. 

\subsection*{Resting-State fMRI data and processing}
We use resting-state fMRI data from the Human Connectome Project (HCP) \cite{smith2013resting} and the Genomics Superstruct Project (GSP) \cite{holmes2015brain} for brain hemisphere classification to study lateralization. Table \ref{tab:data} summarizes the demographic information of the subjects involved in our experiments across both datasets.

\subsubsection*{Human Connectome Project (HCP)}

\paragraph{Acquisition}
All MRI data were collected using the same 3T Siemens Skyra magnetic resonance machines at Washington University in St. Louis with a 32-channel head coil \citesupp{van2013wu}. 
Specifically, rs-fMRI was acquired using a gradient-echo echo-planar imaging (GE-EPI) sequence with the following parameters: repetition time (TR) = 720 ms, echo time (TE) = 33.1 ms, flip angle (FA) = 52°, bandwidth = 2290 Hz/pixel, field of view (FOV) $= 208 \times 180 \text{ mm}^2$, matrix $= 104 \times 90$, voxel size $ = 2 \times 2 \times 2 \text{ mm}^3$, multi-band acceleration factor $= 8$, slices = 72, and total scan time of 1200 frames = 14 min and 24 s \cite{smith2013resting}. During the scan, participants were asked to open their eyes and stare at a white cross on a screen with a black background. There were two rs-fMRI sessions (REST1 and REST2) acquired on two consecutive days, each including two runs with a left-to-right (LR) and a right-to-left (RL) phase encoding direction. The T1-weighted images were acquired by using a magnetized rapid gradient-echo imaging (MPRAGE) sequence with the following parameters: TR = 2,400 ms, TE = 2.14 ms, reversal time (TI) = 1,000 ms, FA = 8°, FOV $ = 224 \times 224 \text{ mm}^2$, voxel size 0.7 $\text{mm}$ isotropic, and total scan time = 7 min and 40 s.

\paragraph{Preprocessing}
The HCP minimal preprocessing pipeline (version 2.0) was utilized, which included steps for magnetic gradient distortion correction, EPI distortion correction, non-brain tissue removal, Montreal Neurological Institute (MNI) standard space registration, and intensity normalization \citesupp{glasser2013minimal}. Resultant data were denoised using independent component analysis (ICA) facilitated by the FIX tool \citesupp{salimi2014automatic}. This tool effectively identifies and eliminates spatiotemporal signal components that originate from non-neuronal or structural noise, with an emphasis on head movement \cite{smith2013resting}. Subsequently, five post-processing steps were utilized based on the minimal preprocessed data, including 1) spatial smoothing with 4 $\text{mm}$ FWHM kernel, twice the voxel resolution of HCP fMRI data \citesupp{fox2005human}; 2) linear detrending to minimize the effects of low-frequency drift \citesupp{lowe1999treatment}; 3) regression of a suite of nuisance variables unrelated to neural signals, such as average signals from white matter (WM) and cerebrospinal fluid (CSF), as well as the whole brain (global signal, GS)\citesupp{hallquist2013nuisance, fox2009global}; 4) bandpass filtering (0.01Hz-0.1Hz) \citesupp{thompson2015frequency,fox2005human}; 5) scrubbing to control effects of transient movement across the time series frames \citesupp{power2012spurious}.  

\subsubsection*{Genomics Superstruct Project (GSP)}
\paragraph{Acquisition}
All imaging data were collected on matched 3T Tim Trio scanners (Siemens Healthcare, Erlangen, Germany) at Harvard University and Massachusetts General Hospital using the vendor-supplied 12-channel phased-array head coil \cite{holmes2015brain}. Structural data included a high-resolution (1.2 $\text{mm}$ isotropic) multi-echo T1-weighted magnetization-prepared gradient-echo image. Functional imaging data were acquired using a GE-EPI sequence sensitive to blood oxygenation level-dependent (BOLD) contrast with the following parameters: TR = 3,000 ms, TE = 30 ms, FA = 85°, voxel size $ = 3 \times 3 \times 3 \text{ mm}^3$, slices = 47, and total scan time of 124 frames = 6 min and 12 s.

\paragraph{Preprocessing}
All fMRI data were preprocessed by SPM \citesupp{friston2003statistical} and GRETNA \citesupp{wang2015gretna} toolkit, including the following steps: 1) removing the first four volumes to assure that the magnetization is at steady state \citesupp{parkes2018evaluation}; 2) slice-timing correction; 3) realignment of all volumes to the first volume to reduce the effects of head motion\citesupp{parkes2018evaluation}; 4) co-registration of GE-EPI data to the native, cropped, high-resolution structural image and then normalizing them to the MNI space through Diffeomorphic Anatomical Registration Through Exponentiated Lie Algebra (DARTEL) algorithm\citesupp{ashburner2007fast}; 5) spatial smoothing with a 6 $mm$ FWHM kernel, twice the voxel resolution of GSP fMRI data \citesupp{fox2005human}; 6) linear detrending to minimize the effects of low-frequency drift \citesupp{lowe1999treatment}; 7) six head motion parameters regression \citesupp{yan2013addressing}, as well as the WM, CSF and GS \citesupp{hallquist2013nuisance, fox2009global}; 8) lowpass filtering (<0.08Hz) \citesupp{yeo2011organization}.

\subsubsection*{Extracting intrahemispheric brain network}
We use intrahemispheric brain network connectivity as features to represent brain hemispheres. Figure \ref{fig:gsda-method}a \ding{172}-\ding{174} illustrates the data processing workflow for obtaining intrahemispheric connections from resting-state time series. Time sequences were extracted using the Brainnetome atlas (BNA) \cite{fan2016human}, which divides the human brain into 246 regions (123 per hemisphere). Pearson correlation was computed to represent the connectivity between brain regions. Following Liang \textit{et al.} \citesupp{liang2014neuroplasticity}, the correlation coefficients were transformed into $z$-scores using Fisher's $z$ transform. For HCP data, we averaged $z$-scores across the RL and LR runs for each session. To extract half-brain features, we reordered the columns and rows of the connectivity matrix to produce two 123 $\times$ 123 matrices, representing the intrahemispheric networks for the two brain hemispheres of each subject. We then extracted the upper (or lower) triangle of these matrices (illustrated as the red and blue areas in \ding{174} of Fig. \ref{fig:gsda-method}) to form two 7503-dimensional feature vectors by BNA for the two hemispheres for experiments.

\begin{table}
    \centering
    \caption{Information of HCP and GSP dataset used for the experiments, where ``M'' denotes male and ``F'' denotes female for sex, ``L'' denotes left-handedness, ``R'' denotes right-handedness, and ``A'' denotes ambidexterity for handedness, and ``SD'' denotes standard deviation.}
    \begin{tabular}{c r r r r r}
        \toprule
         Dataset & \# Total Subjects & Sex (M/F) & Handedness (L/R/A) &  Age (SD) & \# Sessions\\
         \midrule
         HCP \cite{smith2013resting} &  960 & 445/515 & 85/875/0   & 28.7 (3.71) & 2\\
         GSP \cite{holmes2015brain}  & 1570 & 665/905 & 110/1449/11 & 21.5 (2.89) & 1\\
         \bottomrule
    \end{tabular}
    \label{tab:data}
\end{table}

\subsection*{Experimental setting}

\subsubsection*{Multivariate classification algorithm setup}
For all multivariate methods, the classification problem is binary: left brain hemispheres are labeled as 0, and right brain hemispheres are labeled as 1. For GSDA-Logit, sex is utilized as the grouping factor in the experiments, encoding males as 0 and females as 1. Given the binary nature of the grouping factor, the matrix $\mathbf{G}$ simplifies to a vector $\mathbf{g}$ in this experiment. There are two hyperparameters in GSDA-Logit: $\alpha$ and $\lambda$. In this experiment, the value of $\alpha$ is set to 0.1, while $\lambda \in [0, 1.0, 2.0, 5.0, 8.0, 10.0]$. When $\lambda=0$, indicating an absence of the grouping factor dependence for optimizing model weights, GSDA-Logit degenerates to a standard logistic regression for target-group data. A logistic regression classifier implemented in \textit{scikit-learn} \citesupp{scikit-learn} with default hyperparameters is used for learning both first-order multivariate control models and conducting second-order classification.

\subsubsection*{Cross-validation strategy}

\paragraph{First-order classification setting}

Given that each participant provides two brain hemispheres (left and right), the correlation between these hemispheres might impact the effectiveness of machine learning models. To address this, we divided our subjects into two groups: for 50\% of the subjects, we used their left hemispheres for training, and for the remaining 50\%, we used their right hemispheres. The hemispheres not selected for training were then used for testing, as illustrated in \ding{175} of Fig. \ref{fig:gsda-method}a. To further validate our findings, we employed an alternative cross-validation method. This method involves holding 20\% of the subjects as unseen to the models, serving as an additional test set. Training samples are drawn from the remaining 80\% using the same selection process mentioned above. Since HCP data includes two scanning sessions per subject on two different days, the session not involved in training serves as an additional test set. Each cross-validation strategy was iterated 1,000 times, resulting in 1,000 models for each learning task.

\paragraph{Second-order classification setting} 
With the 1,000 first-order models for each task learned, we perform a second-order classification through the following steps:
\begin{enumerate}
    \item Select classification problem of interest, for example, male-specific GSDA models trained on HCP with $\alpha=0.1$, $\lambda=5$ vs . female-specific GSDA models trained on HCP with $\alpha=0.1$, $\lambda=5$, with 1,000 models for each group.
    \item Perform a random stratified split of the 2,000 models into training (80\%) and test (20\%) sets.
    \item Train a standard logistic regression classifier using the scikit-learn \citesupp{scikit-learn} implementation with default hyperparameters on the training set and then evaluate the performance on the test set.
    \item Repeat steps 2 and 3 using different random seeds to split the training and test sets 1,000 times.
\end{enumerate}

\section*{Data availability}
This study used publicly available data from HCP (\url{https://www. humanconnectome.org/}) and GSP (\url{https://www.neuroinfo.org/gsp/}). Original data can be accessed via data use agreements. Processed intrahemispheric network data for classification is available at \url{https://doi.org/10.5281/zenodo.10050233} for HCP and \url{https://doi.org/10.5281/zenodo.10050234} for GSP. The Brainnetome atlas (BNA) atlas is available at \url{http://atlas.brainnetome.org/} and the information about the lobes and gyrus is available at \url{https://pan.cstcloud.cn/web/share.html?hash=6eRCJ0zDTFk}. 

\section*{Code availability}
All custom preprocessing (in Matlab) and analysis (in Python) code can be found in the following repository: \url{https://github.com/shuo-zhou/GSDA-Lateralization}. An online demonstration example is available at: \url{https://colab.research.google.com/github/shuo-zhou/GSDA-Lateralization/blob/main/gsda_demo.ipynb}

\bibliographystylesupp{naturemag-doi}
\bibliographysupp{method_ref}

\section*{Acknowledgements}

This work was supported by the STI 2030-Major Projects (2021ZD0200500, 2021ZD0201701), the National Natural Science Foundation of China (T2325006, 82021004),  and the Fundamental Research Funds for the Central Universities (2233200020).

\section*{Author contributions statement}
S.Z., J.L., H.L., and G.G. conceptualized the study. S.Z. and J.L. designed the research experiments. S.Z. developed the Group-Specific Discriminant Analysis algorithm and performed the classification experiments. J.L. analyzed model weights. Y.J. implemented the preprocessing pipeline and processed the data for experiments. S.Z. and H.W. developed the metric Group-Specificity Index for evaluation. S.Z., J.L., and H.W. wrote the manuscript. H.L. and G.G. performed critical revisions of the article. All authors reviewed and edited the manuscript.

\newpage
\section*{Extended data figure}

\renewcommand\thefigure{Extended Data Fig. \arabic{figure}}    
\setcounter{figure}{0} 

\begin{figure*}[h]
    \includegraphics[width=\textwidth]{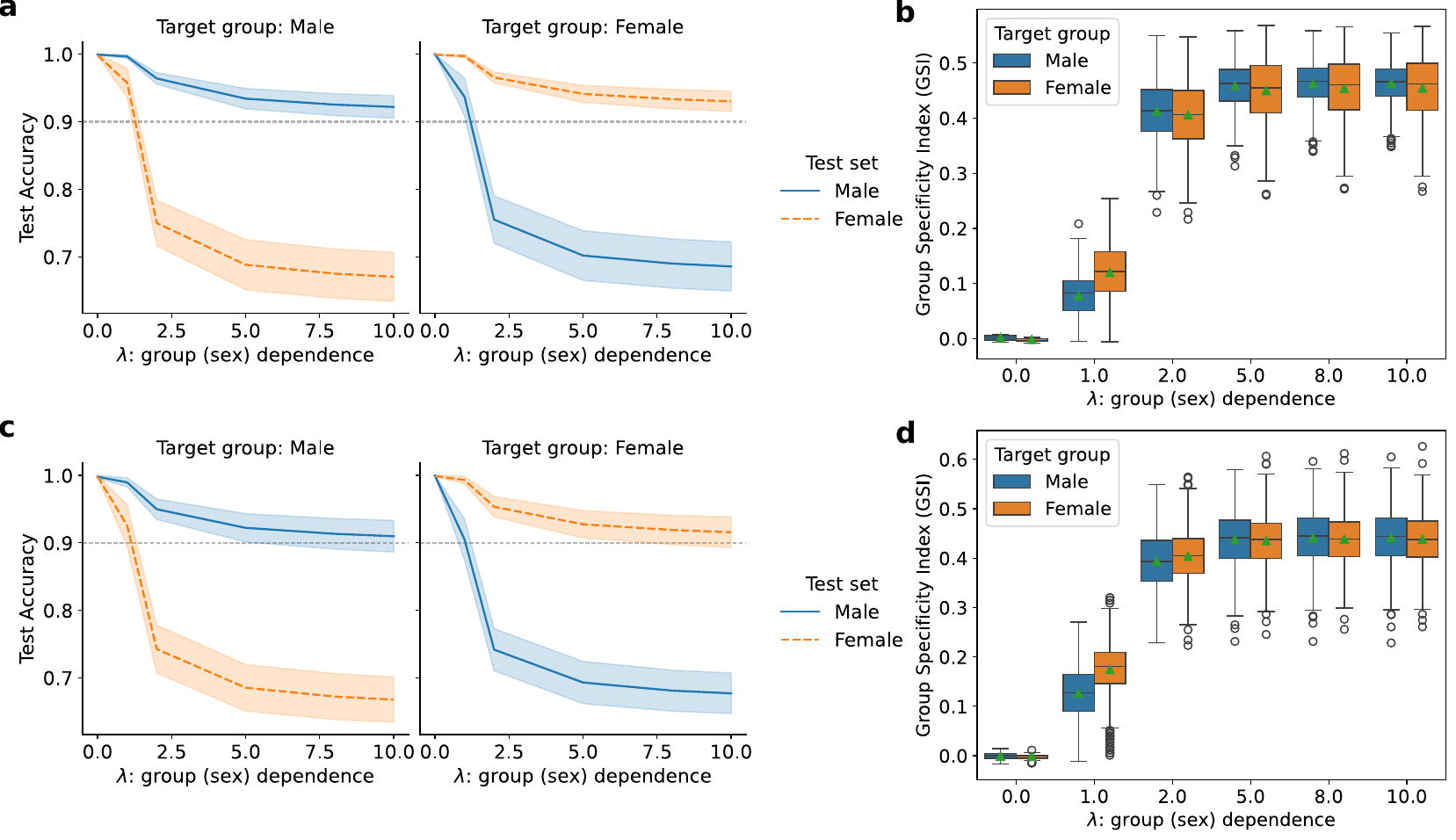}
 \caption{\textbf{Additional results of left vs. right brain classification using GSDA-Logit on HCP data \cite{smith2013resting} with different cross-validation strategies from Fig. \ref{fig:cls_res}}. \textbf{a}, Average test accuracy on the held-out session; for example, training was conducted on the data from the REST1 session, and test was performed on the data from the REST2 session. \textbf{b}, GSI calculated from the test results shown in \ref{fig:hcp_ext_cls_results}a. \textbf{c}, Average test accuracy on the held-out subjects' data; for example, training was conducted on the 80\% subjects' data sampled from the REST1 session, and test was performed on the remaining 20\% subjects' data from REST1 and REST2. \textbf{d}, GSI calculated from the test results shown in \ref{fig:hcp_ext_cls_results}c. The remaining detailed descriptions of the figures, along with the main observations, are the same as those in the caption of Fig. \ref{fig:cls_res}.
 }
 \label{fig:hcp_ext_cls_results}
\end{figure*}

\begin{figure*}[t]
	\includegraphics[width=\textwidth]{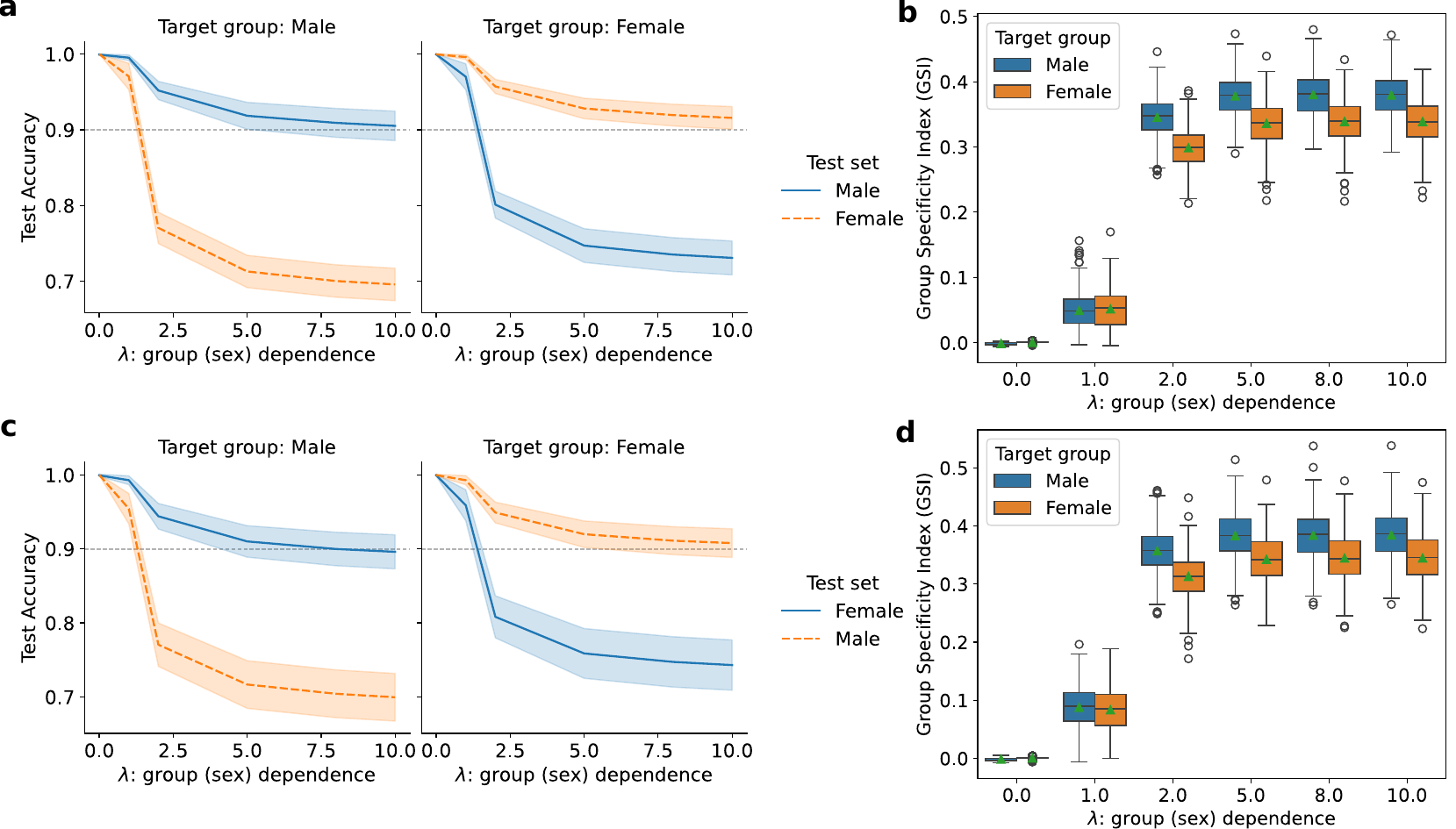}
 \caption{\textbf{Performance of left vs. right brain classification using GSDA-Logit on Brain Genomics Superstruct Project (GSP) data \cite{holmes2015brain}} with respect to the hyperparameter $\lambda$. \textbf{a}, Average test accuracy on the held-out hemispheres, with a cross-validation strategy consistent with that shown in Fig. \ref{fig:cls_res}a. \textbf{b}, GSI calculated from the test results shown in \ref{fig:gsp_cls_results}a. \textbf{c}, Average test accuracy on the held-out subjects' data, with a cross-validation strategy consistent with that shown in \ref{fig:hcp_ext_cls_results}c. \textbf{d}, GSI calculated from the test results shown in \ref{fig:gsp_cls_results}c. The remaining detailed descriptions of the figures, along with the main observations, are the same as those in the caption of Fig. \ref{fig:cls_res}.
 }
 \label{fig:gsp_cls_results}
\end{figure*}

\begin{figure*}
    \includegraphics[width=\textwidth]{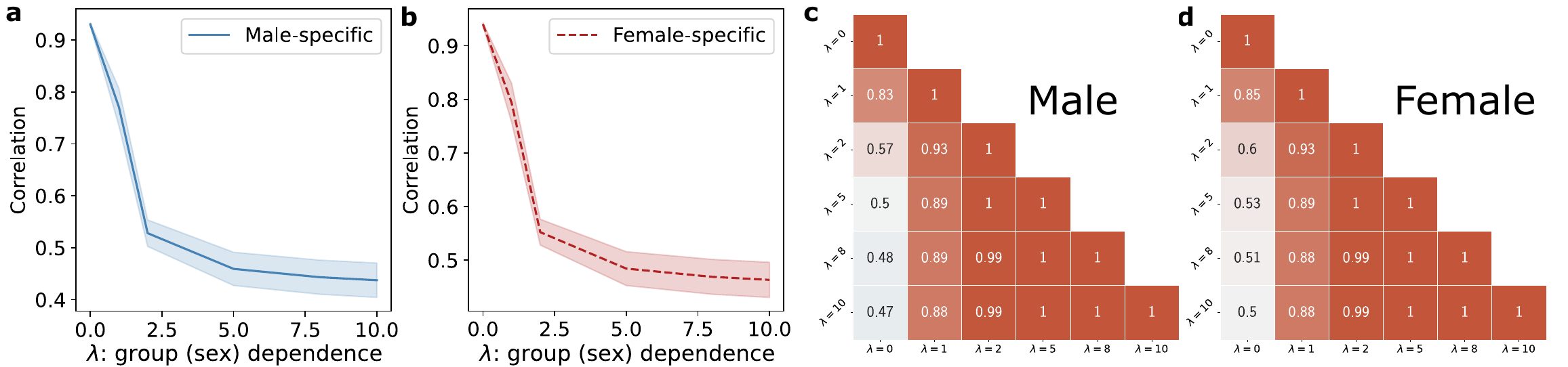}
    \caption{\textbf{Additional Pearson correlation coefficients between model weights learned from GSP data \cite{holmes2015brain}.} \textbf{a}, Correlation between male-specific and multivariate control models. \textbf{b}, Correlation between female-specific and multivariate control models. \textbf{c, d}, Average pairwise correlation for \textbf{c}, male-specific and \textbf{d}, female-specific GSDA models. The main observations are consistent with those in Fig. \ref{fig:main_corr}.}
 \label{fig:gsp_corr}
\end{figure*}

\begin{figure*}
    \centering
    \includegraphics[width=\textwidth]{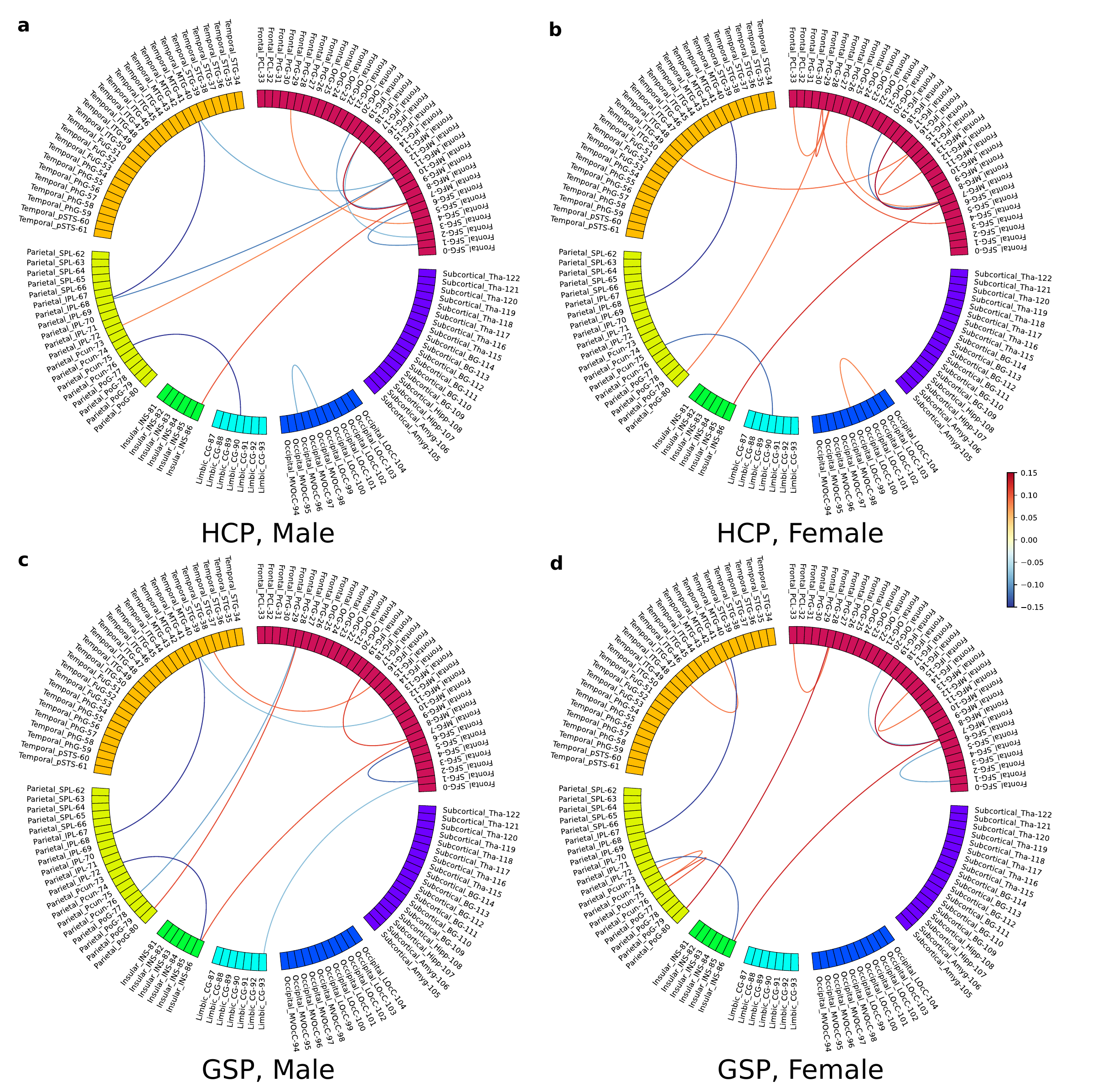}
    \caption{\textbf{Sex-specific lateralized connections} identified by \textbf{a}, male-specific models for HCP (Fig. \ref{fig:edges}a + Fig. \ref{fig:edges}e), \textbf{b}, female-specific models for HCP (Fig. \ref{fig:edges}b + Fig. \ref{fig:edges}f), \textbf{c}, male-specific models for GSP (Fig. \ref{fig:edges}c + Fig. \ref{fig:edges}g), and \textbf{d}, female-specific models for GSP (Fig. \ref{fig:edges}d + Fig. \ref{fig:edges}g).}
 \label{fig:chord-ext}
\end{figure*}

\begin{figure*}
    \centering
    \includegraphics[width=0.88\textwidth]{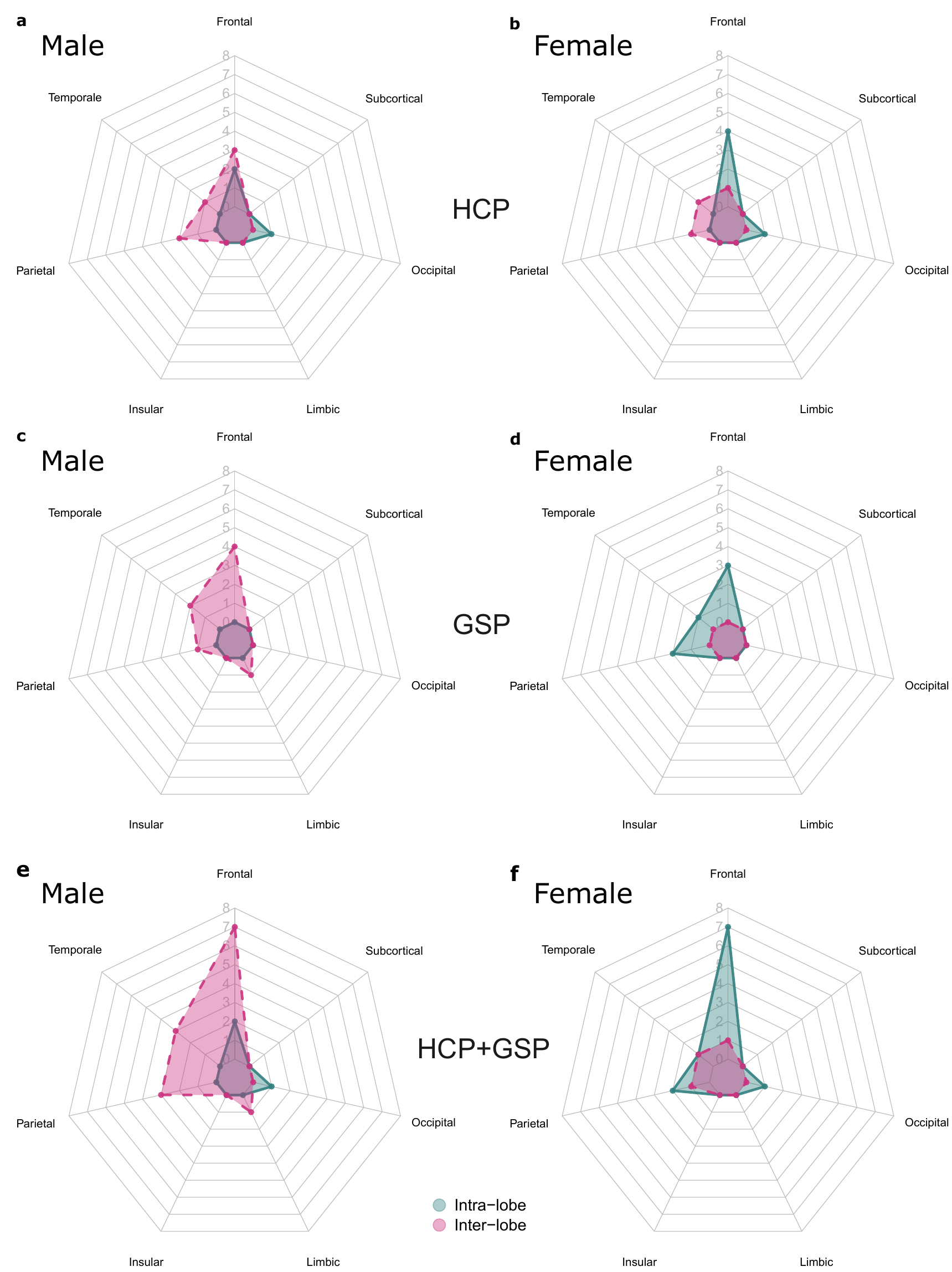}
    \caption{\textbf{Count of the group ``exclusive'' lateralized connections for HCP and GSP (Fig. \ref{fig:edges}e,f,g,h) categorized by associated lobes, and inter- or intra-lobe.} The connections are identified by \textbf{a}, male-specific models for HCP, \textbf{b}, female-specific models for HCP, \textbf{c}, male-specific models for GSP, \textbf{d}, female-specific models for GSP. \textbf{e}, Sum of \ref{fig:rader-ext}a and c. and \textbf{f}, Sum of \ref{fig:rader-ext}b and d.}
 \label{fig:rader-ext}
\end{figure*}

\end{document}